\pdfoutput=1
\documentclass[cernpreprint,atlasdraft=false,coverpage=false,texlive=2016,UKenglish,LANGEDIT=true,LANGSHOW=false,texmf]{atlasdoc}
\LEcontact{Steve Lloyd s.l.lloyd@qmul.ac.uk}
 
\usepackage[]{atlaspackage}
\usepackage{atlasbiblatex}
 
\usepackage{atlasphysics}
 
\graphicspath{{logos/}{figures/}}
 
\usepackage{ANA-STDM-2017-14-PAPER-defs}
\usepackage{atlasjetetmiss}
 
\usepackage{float}
\floatstyle{plaintop}
\restylefloat{table}
 
\usepackage{verbatim}
\usepackage{pgfplots}
\pgfplotsset{width=7cm,compat=1.8}
\usepackage{pgfplotstable}
 
\usepackage{siunitx}
\usepackage{multirow}

\AtlasTitle{Measurement of the jet mass in high transverse momentum $Z(\rightarrow b\overline{b})\gamma$  production at $\sqrt{s}= 13$~\ensuremath{\text{Te\kern -0.1em V}} using the ATLAS detector }
 
\AtlasAbstract{
The integrated fiducial cross-section and unfolded differential jet mass spectrum of high transverse momentum $Z\rightarrow b\overline{b}$\, decays are measured in $Z\gamma$ events in proton--proton collisions at $\sqrt{s} = 13$~\ensuremath{\text{Te\kern -0.1em V}}. The data analysed were collected between 2015 and 2016 with the ATLAS detector at the Large Hadron Collider and correspond to an integrated luminosity of 36.1~fb$^{-1}$. Photons are required to have a transverse momentum  $p_{\mathrm{T}}>175$~\ensuremath{\text{Ge\kern -0.1em V}}. The $Z\rightarrow b\overline{b}$ decay is reconstructed using a jet with $p_{\mathrm{T}} > 200$~\ensuremath{\text{Ge\kern -0.1em V}}, found with the anti-$k_{t}$ $R = 1.0$ jet algorithm, and groomed to remove soft and wide-angle radiation and to mitigate contributions from the underlying event and additional proton--proton collisions. Two different but related measurements are performed using two jet grooming definitions for reconstructing the $Z\rightarrow b\overline{b}$ decay: trimming and soft drop. These algorithms differ in their experimental and phenomenological implications regarding jet mass reconstruction and theoretical precision. To identify $Z$ bosons, $b$-tagged $R = 0.2$ track-jets matched to the groomed large-$R$ calorimeter jet are used as a proxy for the $b$-quarks.  The signal yield is determined from fits of the data-driven background templates to the different jet mass distributions for the two grooming methods. Integrated fiducial cross-sections and unfolded jet mass spectra for each grooming method are compared with leading-order theoretical predictions. The results are found to be in good agreement with Standard Model expectations within the current statistical and systematic uncertainties.
}
 
\author{The ATLAS Collaboration}
 
\AtlasRefCode{STDM-2017-14}
 
\PreprintIdNumber{CERN-EP-2019-115}
 
\AtlasDate{\today}
 
\AtlasJournalRef{Phys. Lett. B 812 (2020) 135991}
\AtlasDOI{10.1016/j.physletb.2020.135991}
\hypersetup{pdftitle={ATLAS document},pdfauthor={The ATLAS Collaboration}}
 
\begin{document}
 
\maketitle
 
\tableofcontents

\section{Introduction}
\label{sec:intro}
% The next lines are included from the .//sections/sec-introduction.tex input file
This Letter presents a measurement of the fiducial and differential jet mass cross-sections of high transverse momentum (\pt) \Z bosons that decay into \bbbar pairs and are produced in association with a photon, denoted by \Zbbgamma. The analysis uses proton--proton (\pp) collision data collected in 2015 and 2016 by the ATLAS detector~\cite{PERF-2007-01} at the Large Hadron Collider (LHC) at a center-of-mass energy of \sqsthirt. This measurement of the unfolded jet mass spectrum of hadronically decaying \Z bosons at the LHC explores the experimental features and phenomenological implications of techniques used to reconstruct boosted bosons -- colour singlets -- decaying into $\bbbar$. Similar measurements of gluons -- colour octets -- decaying into \bbbar pairs have also been made by the ATLAS Collaboration~\cite{STDM-2017-17}. The \Zbbgamma process provides a well-defined experimental signature for measuring massive boosted \Z bosons using high-\pt jets containing pairs of $b$-quarks. A detailed study of the \Zbb signal is important for assessing systematic uncertainties and identification techniques for the measurement of \Hbb in the high-$\pt$ range, as well as for potential \TeV-scale resonances decaying into dibosons, one of them being a \Z boson or a Higgs boson decaying into \bbbar~\cite{Aaboud:2017cxo,Aaboud:2018fgi}.
 
The \Zbbgamma channel offers advantages in accessing the \Zbb signal compared to the inclusive channels studied in Run 1 by ATLAS~\cite{STDM-2013-04} and in Run 2 by CMS~\cite{Sirunyan:2017dgc} since it provides both a useful trigger signature via the photon and an opportunity to directly estimate background processes using the data. Initial results of the modelling of jet kinematics in the \Zbbgamma channel using $13$~\TeV\ data collected by ATLAS are presented in Ref.~\cite{Aad:2019uoz}. The measurement described in this Letter selects \bbbar decays of a \Z boson contained within a single jet, referred to as a \Zjet, with transverse momentum $\ptZjet>200$~\GeV\ and a photon with transverse momentum $\ptg>175$~\GeV. The high-\pt requirement enhances the signal over the dominant \gammajets background production, which has a softer  \pt spectrum. The candidate \Zjet is reconstructed using a `groomed' \antikt~\cite{Cacciari:2008gp} jet with radius parameter $R = 1.0$ (\largeR jet). A multivariate algorithm is used to determine whether $R=0.2$ track-jets that are associated with the \largeR jet are $b$-tagged, i.e.\ if they contain $b$-hadron decay products. The approach to tagging presented in this Letter is built upon a foundation of studies from LHC runs at $\sqrt{s} = 7$ and $8$~\TeV, including extensive studies of jet reconstruction and grooming algorithms~\cite{PERF-2012-02, Krohn2010, Larkoski:2014wba} and detailed investigations of track-jet-based $b$-tagging in boosted topologies~\cite{ATL-PHYS-PUB-2014-013, Aad:2019uoz}.
 
Two different jet grooming algorithms are used to perform the measurement: `trimming'~\cite{Krohn2010}, and `soft drop'~\cite{Dasgupta:2013ihk, Larkoski:2014wba}. The experimental and phenomenological implications for jet mass reconstruction and theoretical precision are different for the two grooming algorithms. The trimming algorithm is the default used in ATLAS to study boosted bosons, chosen as a result of optimisation studies performed from LHC runs at $\sqrt{s}=8$ and 13~\TeV~\cite{PERF-2015-03}. The soft-drop calculations achieve a different theoretical precision and offer advantages such as the formal absence of non-global logarithms. The distribution of the soft-drop mass for QCD processes has now been calculated both at next-to-leading order (NLO) with next-to-leading-logarithm (NLL) accuracy~\cite{Marzani:2017mva, Kang:2018jwa} and at leading order (LO) with next-to-next-to-leading-logarithm (NNLL) accuracy~\cite{Frye:2016okc, Frye:2016aiz}. This level of precision for a jet substructure observable at a hadron collider is surpassed only by the calculation of thrust in $e^+e^-$ interactions~\cite{Becher:2008cf}. Similar calculations are not currently available for trimmed jets.
 
The double differential cross-section of soft-drop jets as a function of the mass and transverse momentum were previously measured by ATLAS~\cite{STDM-2017-04} and CMS~\cite{Sirunyan:2018xdh} in balanced dijet events at $\sqrt{s} = 13$~\TeV. The trimmed jet mass distribution in dijet and $W/Z$+jets events was measured by CMS at $\sqrt{s} = 7$~\TeV~\cite{Chatrchyan:2013vbb}. While previous analyses measured the cross-section of quark and gluon-initiated jets for different grooming algorithms, this analyses measures the mass of \largeR jets containing the hadronic decay products of $Z$ bosons in \Zbbgamma events at $\sqrt{s} = 13$~\TeV.
 
% End of text imported from the .//sections/sec-introduction.tex input file

\section{ATLAS detector}
\label{sec:detector}
% The next lines are included from the .//sections/sec-detector.tex input file
\newcommand{\AtlasCoordFootnote}{ATLAS uses a right-handed coordinate system with its origin at the nominal
interaction point in the centre of the detector. The positive $x$-axis is defined by the direction from the interaction point to the centre of the LHC ring, with the positive $y$-axis pointing upwards, while the beam direction defines the $z$-axis. Cylindrical coordinates $(r,\phi)$ are used in the transverse plane, $\phi$ being the azimuthal angle around the $z$-axis. The pseudorapidity $\eta$ is defined in terms of the polar angle $\theta$ by $\eta=-\ln\tan(\theta/2)$. Rapidity is defined as $y$ = $0.5 \ln[(E + p_z)/(E-p_z)]$ where $E$ denotes the energy and $p_z$ is the component of the momentum along the beam direction. The angular distance $\DeltaR$ is defined as $\sqrt{(\Delta y)^2 + (\Delta \phi)^2}$.}
 
The ATLAS detector at the LHC is a multipurpose particle detector
with a forward--backward symmetric cylindrical geometry and a near $4\pi$ coverage in
solid angle.\footnote{\AtlasCoordFootnote}
It consists of an inner detector (ID) for tracking surrounded by a thin superconducting solenoid
providing a \SI{2}{\tesla} axial magnetic field, electromagnetic and hadronic calorimeters, and a muon spectrometer.
The ID covers the pseudorapidity range $|\eta| < 2.5$.
It consists of silicon pixel, silicon microstrip, and transition radiation tracking detectors.
A new inner pixel layer, the insertable B-layer~\cite{ATLAS-TDR-19, Abbott:2018ikt}, was added at a mean radius of 3.3~cm during the period between Run~1 and Run~2 of the LHC.
Lead/liquid-argon (LAr) sampling calorimeters provide electromagnetic (EM) energy measurements
with high granularity ($|\eta| < 3.2$). The hadronic calorimeter uses a steel/scintillator-tile sampling detector in the central pseudorapidity range ($|\eta| < 1.7$) and a copper/LAr detector in the region $1.5 < |\eta| < 3.2$. The forward regions~($3.2< |\eta| <4.9$) are instrumented with copper/LAr and tungsten/LAr calorimeter modules optimised for electromagnetic and hadronic measurements, respectively.
A muon spectrometer with an air-core toroid magnet system surrounds the calorimeters.
Three layers of high-precision tracking chambers provide coverage in the range $|\eta|<2.7$, while dedicated fast chambers allow triggering in the region $|\eta|<2.4$.
The ATLAS trigger system consists of a hardware-based first-level trigger followed by a software-based high-level trigger~\cite{TRIG-2016-01}.
% End of text imported from the .//sections/sec-detector.tex input file
 
\section{Data and Monte Carlo simulation}
\label{sec:dataMC}
% The next lines are included from the .//sections/sec-dataMC.tex input file
The data were collected in \pp collisions at the LHC with \sqsthirt\ and a 25~ns proton bunch crossing interval during 2015 and 2016. The full data sample corresponds to an integrated luminosity of 36.1~\ifb~after requiring that all detector subsystems were operational during data recording. The uncertainty in the combined 2015--2016 integrated luminosity is 2.1\% \cite{ATLAS-CONF-2019-021}, obtained using the LUCID-2 detector \cite{LUCID2} for the primary luminosity measurements. Collision events were recorded with a trigger selecting events with at least one photon candidate with transverse energy $\et > 140$~\GeV.

Monte Carlo (MC) event samples that include an ATLAS detector simulation~\cite{SOFT-2010-01-bugfix} based on \GEANT4~\cite{Geant4} are used to model the $\Zgamma$ signal and the small $\ttgamma$ and $\Wgamma$ background contributions. In addition, \gammajets MC event samples are used to study the trigger modelling. In addition to the hard scatter, each event was overlaid with additional \pp collisions (pile-up) according to the distribution of the average number of \pp interactions per bunch crossing, $\avgmu$, observed in data.
These additional \pp collisions were generated with \PYTHIAV{8.1}~\cite{pythia8Gen} using the ATLAS A2 set of tuned parameters~\cite{ATL-PHYS-PUB-2012-003} and the NNPDF23LO~\cite{Ball:2012cx} parton distribution function (PDF) set. Simulated events were then reconstructed with the same algorithms as those run on collision data.
 
The $\Zgamma$ signal was modelled using the LO \SHERPAV{2.1.1}~\cite{Gleisberg:2008ta} generator, with the CT10 NLO~\cite{Lai:2010vv} PDF set; the sample is flavour inclusive ($Z(\rightarrow qq)\gamma$). An alternative $\Zgamma$ sample was produced with \MADGRAPHV{5.2}~\cite{Alwall:2014hca}, which generated LO matrix elements that were then parton showered with \PYTHIAV{8.1} using the NNPDF23LO PDF set and the ATLAS A14 set of tuned parameters~\cite{ATL-PHYS-PUB-2014-021} for the underlying event. This alternative signal sample is used to determine the systematic uncertainty associated with the signal modelling.
 
The \gammajets samples were also generated with \SHERPAV{2.1.1} and the CT10 NLO PDF set. The matrix element was configured to allow a photon with up to three partons in the final state. The $\ttgamma$ processes were modelled by \MADGRAPHV{5.2} interfaced to \PYTHIAV{8.1}. NLO corrections were applied to the $\ttgamma$ cross-section~\cite{Aaboud:2018hip}. The $\Wgamma$ MC samples with hadronically decaying \Wboson bosons were generated using \SHERPAV{2.1.1}, with a configuration similar to that used for the $\Zgamma$ sample. Predictions for $\Wgamma$ production were normalised according to the cross-sections provided by the generator.
 
% End of text imported from the .//sections/sec-dataMC.tex input file
 
\section{Event reconstruction and selection}
\label{sec:reconstruction}
% The next lines are included from the .//sections/sec-reconstruction.tex input file
Events are required to have a reconstructed primary vertex, defined as the vertex with at least two reconstructed tracks with $\pt>0.4$~\GeV~and with the highest sum of squared transverse momenta of associated tracks~\cite{ATL-PHYS-PUB-2015-026}. 
 
Hadronically decaying high-\pt \Zbb candidates are identified using \largeR jets to capture both $b$-quarks, since they will be very close due to the high Lorentz boost. The two different jet grooming algorithms considered in the analysis, trimming and soft drop, differ in their \pileup mitigation and mass resolution performance.

\paragraph{Trimmed calorimeter jets}
Trimmed calorimeter jets are reconstructed from noise-suppressed topological clusters (topoclusters) of calorimeter energy deposits calibrated to the local hadronic scale (LC)~\cite{PERF-2014-07}, using the \antikt\ algorithm with radius parameter $R=1.0$ implemented in \Fastjet~\cite{Cacciari200657, Cacciari:2011ma}. Trimmed calorimeter jets are those jets to which the trimming algorithm~\cite{Krohn2010} is applied. The aim of this algorithm is to improve the jet mass resolution and its stability with respect to \pileup by discarding the softer components of jets that originate from initial-state radiation, \pileup interactions, or the underlying event. This is done by reclustering the constituents  of the initial \largeR jet, using the \kt\ algorithm~\cite{Catani:1993hr,Ellis:1993tq}, into subjets with radius parameter $\Rsub=0.2$ and removing any subjet that has a \pT\ less than 5\% (\fcut) of the parent jet \pT. The jet mass \mjet, the main observable in this analysis, is defined as the magnitude of the four-momentum sum of constituents inside a jet. It is referred to as the calorimeter-based mass if it is calculated using the topoclusters as constituents, or as the track-assisted jet mass~\cite{Aaboud:2018kfi} if it is estimated by using tracking information. The jet mass for trimmed jets is defined as the weighted combination of the calorimeter-based mass and the track-assisted jet mass~\cite{Aaboud:2018kfi}, where each input mass is weighted by a factor proportional to their inverse-squared mass resolution.

\paragraph{Soft-drop calorimeter jets}
Soft-drop calorimeter jets are formed by the application of the soft-drop algorithm~\cite{Larkoski:2014wba} to the \antikt\ $R = 1.0$ jets described above, with additional topological cluster preprocessing that is described below. The soft-drop algorithm is designed to remove soft and wide-angle radiation and also contamination from \pileup. In the first step of the grooming algorithm, the \antikt\ $R = 1.0$ jets are reclustered with the Cambridge--Aachen (C/A)~\cite{Dokshitzer:1997in,Wobisch:1998wt} algorithm so that the constituents are combined purely according to their angular separation. The soft-drop algorithm then reverses the C/A algorithm clustering history and removes the softer subjet at a specific step of the C/A clustering history unless the soft-drop condition is fulfilled:
\begin{equation*}
\frac{\mathrm{min}(p_\textrm{T1}, p_{\mathrm{T2}})}{p_{\mathrm{T1}} + p_{\mathrm{T2}}} > \zcut \left( \frac{\Delta R_{12}}{R_0}\right) ^\beta\ ,
\end{equation*}
where \zcut\ and $\beta$ are algorithm parameters, $p_\textrm{T1}$ and $p_\textrm{T2}$ are the transverse momenta of the declustered subjets at each history step, $\Delta R_{12}$ is the distance between the subjets in the ($\eta,\phi$) space and $R_0$ is a threshold corresponding to the jet radius.  The parameters $\beta = 0$ and $\zcut = 0.1$ are used in the analysis, based on the studies in Ref.~\cite{ATL-PHYS-PUB-2017-020}. The final measurement is performed for jet mass $\massjet > 30~\GeV$, which implies that any collinear divergence is regulated and the measurement remains protected against collinear singularities. The soft-drop jet mass exhibits a \pileup dependence with the chosen parameters and therefore a special version of \pileup suppressed topological clusters are used to construct the jets that are then groomed with the soft-drop algorithm.  Specifically, the SoftKiller (SK) algorithm~\cite{Cacciari:2014gra} is used in conjunction with Constituent Subtraction (CS)~\cite{Berta:2014eza,ATLAS-CONF-2017-065} based on the studies presented in Ref.~\cite{ATL-PHYS-PUB-2017-020}. CS is applied before the SK algorithm. The CS is an extension of the \pileup subtraction based on jet area~\cite{Cacciari:2007fd}. The algorithm proceeds as follows. First, virtual particles with infinitesimally small $\pt$ (ghosts) are added to the event (each covering a fixed area in the $\eta$--$\phi$ plane) with energy density matching the median energy density of the event. Second, the added ghosts are matched to the topological clusters in $\eta$--$\phi$ space and only those within $\Delta R = 0.25$ of the topocluster are further considered for the \pileup removal procedure. The algorithm proceeds then iteratively through each topocluster--ghost pair in order of ascending $\Delta R$. If the $\pt$ of the topocluster is larger than that of the matched ghost, the $\pt$ of the individual topocluster is corrected by subtracting the $\pt$ of the ghost and the ghost is removed. Otherwise the $\pt$ of the topocluster is subtracted from the $\pt$ of the ghost and the $\pt$ of the topocluster is set to zero.
The SK algorithm exploits the characteristic that particles originating from \pileup collisions are softer than those from the hard-scattering collision and removes particles that fall below a certain \pT threshold, determined on an event-by-event basis. The \pileup suppressed topological clusters after CS and SK are used as input to the soft-drop jet reconstruction.  The calorimeter-based jet mass is used for soft-drop jets.

\paragraph{All groomed jets}
A dedicated MC-based calibration, similar to the procedure used in Ref.~\cite{Aaboud:2018kfi}, is applied to correct the jet \pt and mass of both the trimmed jets and the soft-drop jets to the particle level. To account for semileptonic decays of the $b$-hadrons, the four-momentum of the closest reconstructed muon candidate within $\Delta R=0.2$ of the $b$-tagged track-jet is taken into account in the calorimeter-based component of the jet mass observable (see below for the description of the track-jet definition and $b$-tagging). Muon candidates are identified by matching ID tracks to full tracks or track segments reconstructed in the muon spectrometer. Muons are required to have $\pt > 10$~\GeV\ and $|\eta|<2.4$, and to satisfy the \textit{loose} identification criteria of Ref.~\cite{PERF-2015-10}, which impose quality requirements on the tracks, but no isolation criteria are applied. A calibration is applied to correct the muon transverse momentum, and reconstruction and identification efficiency scale factors, derived from $J/\psi \rightarrow \mu^+\mu^-$ and $Z\rightarrow \mu^+\mu^-$ events~\cite{PERF-2015-10}, are applied to simulation. \LargeR\ jets are required to  have $\pt > 200$~\GeV\ and $|\eta| < 2.0$.  A comparison of the calibrated \Zjet mass distribution and the particle-level jet mass distribution for trimmed jets and soft-drop jets is shown in \Fig{\ref{fig:jetmass}}. Particle-level jets, used in the unfolding procedure described in Section~\ref{sec:unfolding}, are built from stable final-state particles (defined as those with proper lifetime $\tau$ corresponding to $c\tau>10$~mm) excluding muons and neutrinos and using the same jet reconstruction algorithms used for calorimeter jets. Similarly to the muon-in-jet correction at reconstruction level described in \Sect{\ref{sec:reconstruction}}, particle-level muons are added to the particle-level jet if they are within $\DeltaR = 0.2$ of a $b$-hadron. The mass of particle-level jets is defined as the invariant mass of the four-vector sum of its constituents. The jet mass distribution of soft-drop jets is significantly broader than that of trimmed jets for both the reconstructed jet mass, \Fig{\ref{fig:jetmass-reco}}, and the particle-level jet mass, \Fig{\ref{fig:jetmass-truth}}, whereas the distribution for trimmed jets is more asymmetric than for soft-drop jets at particle-level. This asymmetry is a feature of the trimming algorithm and is independent of the quark flavour from the $Z$ boson decay. The jet mass is very stable with respect to pile-up for both jet definitions~\cite{Aad:2020elc}. The soft-drop jets exhibit basically no pile-up dependence due to the constituent-level pile-up suppression techniques while the trimmed jet mass varies by 0.14~\GeV\ per reconstructed vertex~\cite{Aad:2020elc}. While Ref.~\cite{Aad:2020elc} focusses on the hadronic decay of $W$ bosons, it was found that the same conclusions hold as well for jets containing heavy flavour decays of $Z$ bosons.

\begin{figure}[htbp!]
\begin{center}
\subfloat[]{\includegraphics[width = 0.49\textwidth]{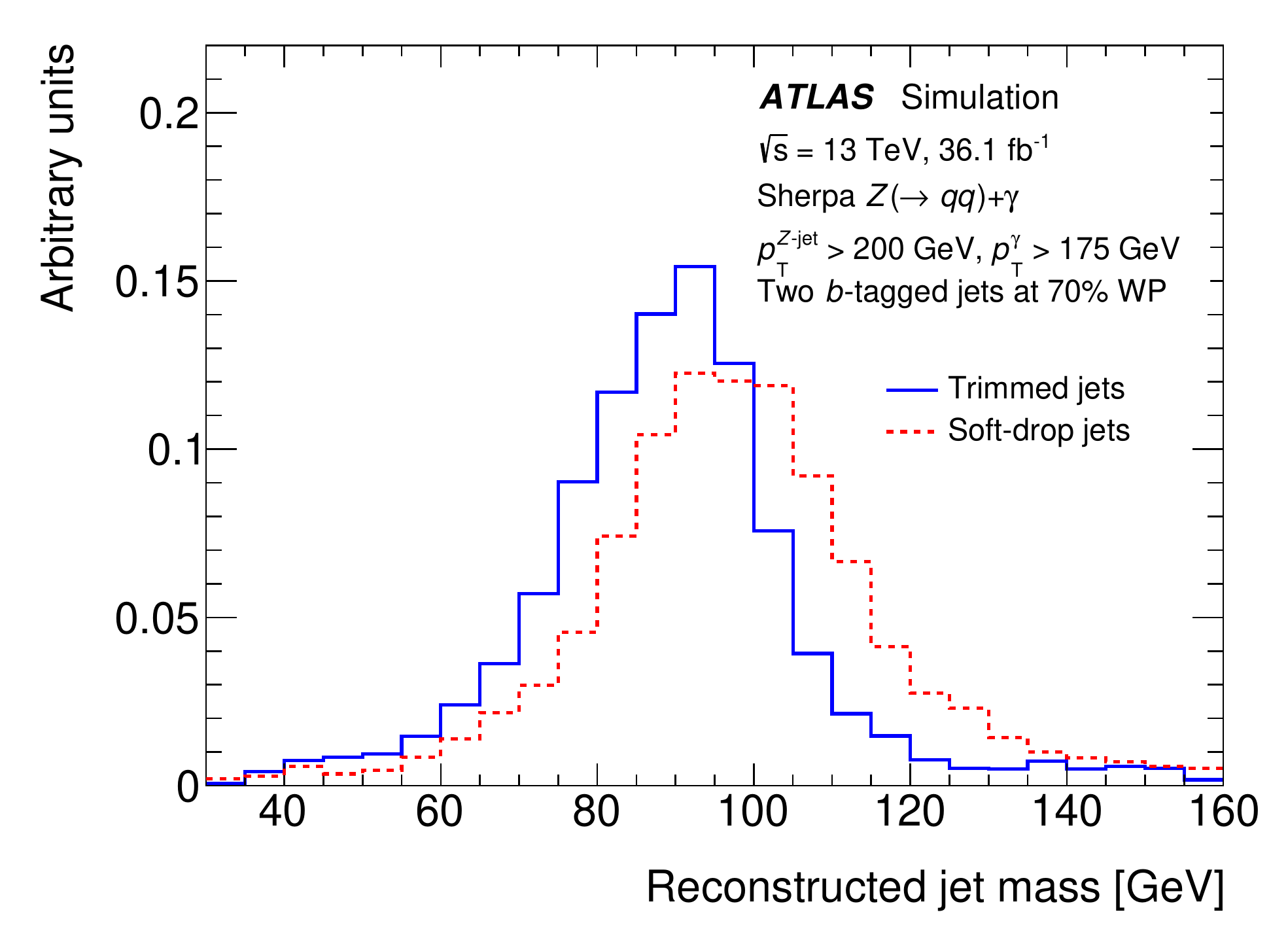}\label{fig:jetmass-reco}}
\subfloat[]{\includegraphics[width = 0.49\textwidth]{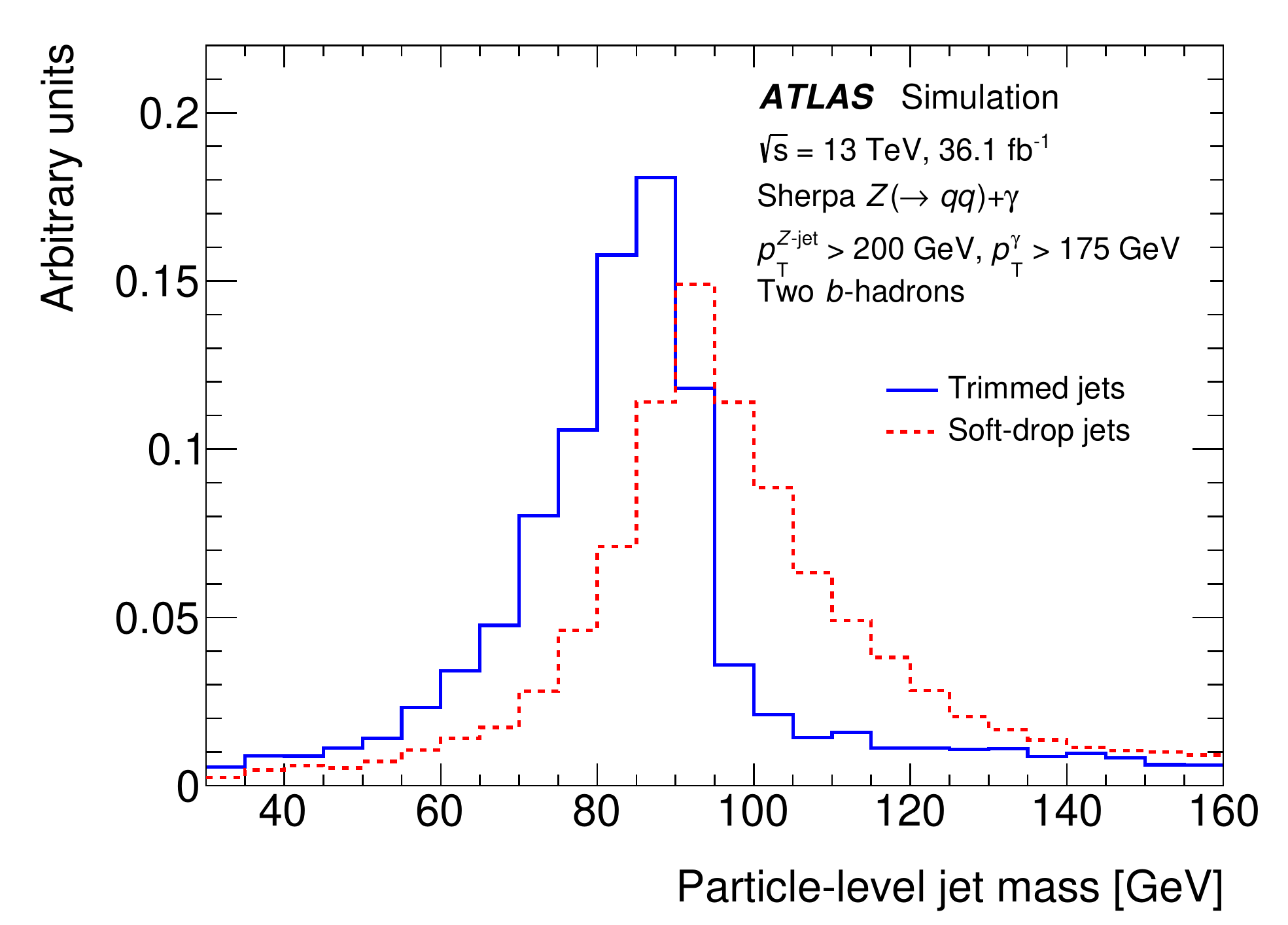}\label{fig:jetmass-truth}}
\end{center}
\caption{Comparison of (a) the calibrated reconstructed \Zjet mass distribution and (b) the particle-level jet mass distribution of soft-drop (dashed line) and trimmed jets (solid line) in the signal region in the \Zgamma sample.}
\label{fig:jetmass}
\end{figure}
 
\paragraph{Track-jets} Small-radius jets formed from charged-particle tracks are used as probes of $b$-hadrons associated with \largeR jets that may contain the candidate \Zbb jets. Track-jets are built with the \antikt\ algorithm with a radius parameter of $R=0.2$~\cite{ATL-PHYS-PUB-2014-013} from at least two ID tracks with $\pt>0.5$~\GeV\ and $|\eta| < 2.5$~\cite{PERF-2016-05}. Only track-jets with $\pt>10$~\GeV\ and $|\eta|<2.5$ are used and they are associated with the \largeR calorimeter jets via ghost-association~\cite{Cacciari:2008gn,Cacciari:2007fd}, in which the track-jets are included in the jet clustering procedure with infinitesimally small $\pt$ such that they have no effect on the jet clustering result. Track-jets containing $b$-hadron decay products are tagged with a multivariate algorithm known as MV2c10, which exploits the presence of large-impact-parameter tracks, the topological decay chain reconstruction and the corresponding displaced vertices from $b$-hadron decays~\cite{PERF-2012-04,ATL-PHYS-PUB-2016-012}.  The MV2c10 algorithm is configured to achieve an efficiency of 70\% for tagging $b$-jets in a MC sample of $\ttbar$ events, while rejecting 80\% of $c$-jets and more than 99\% of light (quark or gluon) jets in the same sample. This configuration is referred to as the 70\% working point (WP). For MC samples, the tagging efficiencies are corrected to match those measured in data~\cite{PERF-2016-05,ATLAS-CONF-2018-006,ATLAS-CONF-2018-001}. These small-radius track-jets are referred to as $b$-jets. By using this small-$R$ definition, $b$-jets can be reliably identified in the dense environment of boosted bosons. Consequently, the number of associated $b$-jets (\nbjet) provides an essential criterion for the identification of merged $\Zbb$ decays.
 
\paragraph{Photons} Photon candidates are reconstructed from clusters of energy deposits in the EM calorimeter~\cite{PERF-2017-02}. The photon energy is calibrated by applying the energy scales measured with $Z\rightarrow e^+e^-$ decays~\cite{PERF-2017-03}. Identification requirements are applied to reduce the contamination from $\pi^0$ or other neutral hadrons decaying into photons. Requirements on the shower shape in the EM calorimeter and on the energy fraction measured in the hadronic calorimeter are used to identify photons. Photons must satisfy the \textit{tight} identification and isolation criteria defined in Ref.~\cite{PERF-2017-02}, and must have  $|\eta|<1.37$ or $1.52<|\eta|<2.37$. For MC samples, the photon reconstruction, identification and isolation efficiencies are corrected to match those measured in data~\cite{PERF-2017-02,PERF-2017-03}. The selected photon is required to have $\ptg>175$~\GeV, which is determined by an optimisation of the expected signal significance~\footnote{The expected signal significance is defined as $s/\sqrt{s+b}$, where $s$ is the number of expected signal events and $b$ is the number of expected background events.}, and to ensure that the trigger is fully efficient. The efficiency of the photon selection ranges between 95\% and 98\% for photons with $\ptg > 175$~\GeV\ depending on the pseudorapidity of the photon. These selection criteria are inverted to form a sample of \textit{non-tight} photons for the background estimate described in \Sect{\ref{sec:background}}.
 
Quality requirements are applied to photon candidates to identify those arising from instrumental problems or non-collision background~\cite{LARG-2013-01}, and events containing such candidates are rejected. In addition, quality requirements are applied to remove events containing spurious jets from detector noise or out-of-time energy deposits in the calorimeter from cosmic rays or other non-collision sources~\cite{ATLAS-CONF-2015-029}.
 
Selected events are required to have at least one groomed \largeR jet and at least one photon, with $\Delta R(\mathrm{jet}, \gamma)>1.0$ from the groomed \largeR jet axis. The groomed \largeR jet is required to have $\ptZjet>200$~\GeV\ to capture both of the decay products of the \Zbb decay, i.e.\ both jets from the $b$-quarks should be fully contained in the groomed \largeR jet. 
In the signal region, the jets identified as candidate \Zbb decays must contain at least two ghost-associated track-jets, and the two with the highest \pt must be tagged as $b$-jets ($\nbjet=2$).

% End of text imported from the .//sections/sec-reconstruction.tex input file

\section{Signal and background estimation}
\label{sec:background}
% The next lines are included from the .//sections/sec-background.tex input file
To extract the \Zbbgamma signal from the data, signal and background templates obtained from MC simulation and from data are fitted to the observed \Zbb candidate jet mass distribution using a binned maximum-likelihood fit. This procedure is repeated separately for each of the groomed jet definitions used to perform the measurement. The dominant background is \gammajets with gluon to $\bbbar$ splitting. Less significant background contributions are due to \ttgamma and \Wgamma processes. Other backgrounds such as multijet and $W/Z$+jet processes, where a jet is misidentified as a photon, and the associated production of a Higgs boson with a $\gamma$ are found to be negligible ($<1$\%).
 
Templates of the jet mass distribution for the \Zbbgamma signal, and for the \ttgamma and \Wgamma backgrounds, are determined from MC simulation. In contrast, the template used to estimate the dominant background contribution from \gammajets processes is derived directly from the measured data without input from MC simulation. 
It is especially important to minimise the reliance on MC simulations for this process, as MC generators have not been tested thoroughly in the high \pt region of the $\bbbar$ production phase space for \gammajets.
 
The data-driven background estimate of the jet mass distribution for the \gammajets process relies on two features of the final state: the $b$-jet multiplicity (i.e.\ \nbjet) and the photon identification criteria (i.e.\ tight vs non-tight). The $b$-jet multiplicity requirement is used to isolate the \gammajets process, which dominates in samples with $\nbjet=0$ or 1. Furthermore, the ratio of \gammajets yields (\Ngammajets) in events with tight compared with non-tight photons is observed to be approximately independent of \nbjet. These two characteristics are used to model the expected \gammajets yield in the signal region via a transfer-factor (TF) method. This method extrapolates the signal region (SR) yield from control regions (CRs) with $\nbjet=1$ and the shape of the jet mass distribution in the signal region for the \gammajets background from CRs with $\nbjet=2$ but non-tight photons. The definitions of the different CRs are summarised in \Tab{\ref{tab:CRSR-defs}}. In these CRs, the \ttgamma and \Wgamma contributions are subtracted from the data, as the mass shape differs from that of \gammajets.
 
% The next lines are included from the .//tables/table-CRSR-definitions.tex input file
\begin{table}[h]
\begin{center}
\begin{tabular}{l | c c c}
\toprule
& $\nbjet=0$ & $\nbjet=1$ & $\nbjet=2$ \\
\midrule
Non-tight $\gamma$  & CR-A       & CR-C       & CR-E       \\
Tight $\gamma$      & CR-B       & CR-D       & SR         \\
\bottomrule
\end{tabular}
\caption{Definitions of the control regions (CR) and the signal region (SR) used for the data-driven background estimate of the \Gammajets process.}
\label{tab:CRSR-defs}
\end{center}
\end{table}
% End of text imported from the .//tables/table-CRSR-definitions.tex input file
 
The \gammajets background estimates are constructed in 10~\GeV\ bins of \Zbb candidate jet mass. A bin size of 10 GeV is chosen based on the \largeR jet mass resolution. For each jet mass bin, $i$, in each CR, the estimated yield of \gammajets events in that bin is calculated as:
\begin{equation*}
\NCRigjets = \NCRi-\NCRittgamma-\NCRiWgamma
\label{eq:CR-gjets-yield}
\end{equation*}
where $\NCRittgamma$ and $\NCRiWgamma$ are the number of \ttgamma and \Wgamma events, respectively, taken directly from the MC simulation. 
The systematic uncertainties for \ttgamma and \Wgamma contributions are described in \Sect{\ref{sec:systematics}}. The contribution from signal events in each of these control regions is negligible compared to other processes and has no impact on the background estimation. 
 
To obtain the estimate of the number of \gammajets events present in each bin of the jet mass distribution in the SR (\NSRigjets), the jet mass distribution from CR-E (\NCREigjets) is multiplied by a TF determined from the $\nbjet=1$ regions: CR-C and CR-D. This procedure may be summarised as
\begin{equation*}
\NSRigjets = \left(\frac{\NCRDigjets}{\NCRCigjets}\right) \NCREigjets \, ,
\label{eq:SR-yield}
\end{equation*}
where the ratio $\NCRDigjets/\NCRCigjets$ is the TF in each bin of the jet mass distribution. The value of the TF varies with jet mass, ranging from 1.2 for jet masses of 30~\GeV\ to 0.8 at 160~\GeV, and is within 5\% of unity from 50 to 110~\GeV. With this TF method, the shape of the jet mass distribution in the signal region is determined from CR-E (with $\nbjet=2$) and the normalisation of each bin is determined from the $\nbjet=1$ control regions.
 
The validity of this approach relies on the assumption that the TF does not depend on \nbjet. This is tested in data, using the $\nbjet=0$ sample as a cross-check, and in MC simulation using $\nbjet=0,1,$ and 2. The differences in the TFs between the $\nbjet=0$ and $\nbjet=1$ control regions in data are taken as systematic uncertainties in the TFs, as described in \Sect{\ref{sec:systematics}}. The TF approach was further validated by comparing the TF, determined from the $\nbjet=1$ regions, in data and MC. Within their statistical precision, the TF were found to be compatible. Additionally, the reconstructed jet mass distribution of the data-driven background estimate in the SR is in good agreement with the jet mass distribution in \gammajets events simulated with \SHERPAV{2.1.1}.

% End of text imported from the .//sections/sec-background.tex input file
 
\section{Definition of the observable and correction for detector effects}
\label{sec:unfolding}
% The next lines are included from the .//sections/sec-unfolding.tex input file
The reconstructed jet mass distributions from the signal regions are corrected to particle level in order to measure the full differential cross-section of the \Zbb jet mass. Unfolding accounts for the effects of detector resolution and inefficiency and allows direct comparisons with particle-level predictions. The particle-level event selection is similar to the selection described in \Sect{\ref{sec:reconstruction}}. 
Events are required to have at least one particle-level jet with $\pt>200~\GeV$, $|\eta|<2.0$ and two ghost-associated $b$-hadrons. They are also required to have a particle-level photon with $\pt>175~\GeV$,  $|\eta|<1.37$ or $1.52<|\eta|<2.37$ and $\Delta R(\mathrm{jet}, \gamma)>1.0$.

The estimated background jet mass spectrum is subtracted from the data in the signal region, as discussed in \Sect{\ref{sec:results}}. The background-subtracted distribution of the reconstructed jet mass is then unfolded using an iterative Bayesian technique~\cite{2010arXiv1010.0632D} with one iteration. This technique is implemented in the RooUnfold framework~\cite{Adye:2011gm}. One iteration is chosen to minimise the statistical uncertainties as well as the uncertainties associated with the unfolding method estimated with a data-driven closure test as described in \Sect{\ref{sec:systematics}}. The unfolding procedure corrects for bin migrations between the particle-level and the reconstructed jet mass distribution using a response matrix that describes the probability for an event with a particle-level jet mass in bin $i$ to be reconstructed in bin $j$. The response matrix is constructed from events that satisfy the event selection and fiducial region criteria at both the particle level and the reconstruction level. The particle-level jets and reconstructed jets are required to be matched within $\DeltaR = 0.75$. Furthermore, the unfolding procedure corrects for events that satisfy either the particle-level or reconstructed selection criteria, but not both. The response matrix is obtained from the \Sherpa \Zbbgamma signal MC simulation. The \Zbb candidate jet mass distribution was rebinned in the high jet mass region to improve the correlation between the reconstructed and particle-level jet mass. The unfolding results are found to be compatible when increasing the number of iterations used, at the expense of an increase in the statistical uncertainties. The unfolding procedure is also validated by unfolding the jet mass spectra using a singular value decomposition (SVD) technique~\cite{Hocker:1995kb} and within the statistical precision of the measurements, the results are found to be compatible.

% End of text imported from the .//sections/sec-unfolding.tex input file
 
\section{Systematic uncertainties}
\label{sec:systematics}
% The next lines are included from the .//sections/sec-systematics.tex input file
Various sources of systematic uncertainties impact the \Zbb candidate jet mass distribution. These are classified into experimental and theoretical uncertainties, and uncertainties related to the background estimate and the unfolding procedure. The systematic uncertainties can have an impact on the shape of the jet mass distribution and on the signal and background yields. Systematic uncertainties are evaluated by varying each source by plus or minus one standard deviation of its uncertainty. The fit is repeated for each variation and the jet mass distribution unfolded to particle level. The jet energy and mass scale uncertainties are treated as correlated while all other sources of systematic uncertainties are treated as uncorrelated.
 
The impact of the systematic uncertainties on the integrated fiducial cross-section measurements, grouped by source, is summarised in \Tab{\ref{tab:SysTable_crossSection_smoothed}}.
 
For groomed large-$R$ jets, the uncertainties in the energy and mass scales are estimated by using the double-ratio technique described in Ref.~\cite{Aaboud:2018kfi} by comparing the calorimeter jet properties with the measurements of the same jet reconstructed from tracks in the ID. The uncertainties in the jet mass and energy resolutions are assessed by applying additional smearing of the jet observables according to the uncertainty in their resolution measurements. An absolute uncertainty of 2\% is used for the jet energy resolution while a relative uncertainty of 20\% is used for the jet mass resolution, consistent with previous studies of both the trimmed and soft-drop jet definitions~\cite{STDM-2011-19,STDM-2017-04}.

The $b$-tagging uncertainty is evaluated by varying the data-to-MC corrections in various kinematic regions, based on the measured tagging efficiency and mistag rates. These variations are applied separately to $b$-hadron jets, $c$-hadron jets, and light jets, leading to three uncorrelated systematic uncertainties. An additional uncertainty is included to account for the extrapolation to jets with \pt\ beyond the kinematic reach of the data calibration~\cite{PERF-2016-05,ATLAS-CONF-2018-006,ATLAS-CONF-2018-001}.
 
The impact of the systematic uncertainties on the photon reconstruction, identification and isolation efficiencies is studied by varying the scale factors, used to correct the respective efficiencies in simulation to match those observed in data, within their uncertainties. The uncertainties are determined from data samples of $Z\rightarrow \ell^+\ell^-\gamma$ (with $\ell=e,\mu$), $Z\rightarrow e^+e^-$, and inclusive photon events, using the methods described in Ref.~\cite{PERF-2017-02}. Uncertainties in the photon energy scale and resolution are also taken into account~\cite{PERF-2017-03}.
 
The uncertainties associated with the muon momentum calibration and resolution, and the reconstruction and identification efficiency scale factors, are derived from $Z\rightarrow \mu^+\mu^-$ events~\cite{PERF-2015-10}.
 
The uncertainty associated with the modelling of \pileup in the simulation is assessed by varying the reweighting of the \pileup in the simulation within its uncertainties. This uncertainty covers the difference between the ratios of predicted and measured inelastic cross-section values~\cite{STDM-2015-05}.
 
The efficiency of the photon trigger is 100\% for photons with $\ET^{\gamma} > 175$~\GeV, with an uncertainty of 0.5\% that is propagated through the unfolding.
 
The systematic uncertainties associated with the data-driven background template are estimated by deriving the bin-by-bin normalisation from CR-A and CR-B with $\nbjet=0$ instead of from the $\nbjet=1$ CRs as described in \Sect{\ref{sec:background}} (referred to as 'Transfer factor: 0-tag vs 1-tag'). An additional uncertainty in the bin-by-bin normalisation of the background template is derived by varying the jet mass distributions in CR-C and CR-D (with $\nbjet=1$) within their statistical uncertainty (referred to as 'Transfer factor: statistical').
 
The signal and background yields are estimated by performing a simultaneous fit to the data. The uncertainty in the normalisation of the background template, arising from the statistical uncertainty in the data, is referred to as the fit statistical uncertainty in the following.

The modelling uncertainties affecting the \Wgamma process are derived by comparing the nominal \SHERPAV{2.1.1} sample with one produced using the \MADGRAPH~\cite{Alwall:2014hca} generator interfaced to \PYTHIAV{8}. For the \ttgamma background, three different sources of modelling uncertainties are considered: the uncertainty due to the parton shower and hadronisation is estimated by comparing the nominal samples produced using \MADGRAPH interfaced to \PYTHIAV{8}, with \MADGRAPH interfaced to \HERWIGV{7}~\cite{Herwigpp,Bellm:2015jjp}; the uncertainty due to different initial- and final-state radiation conditions is estimated by using \PYTHIAV{8} tuned parameters with high or low QCD radiation activity;
and the uncertainties due to the choice of renormalisation and factorisation scales are estimated by using alternative samples with the scales varied independently by factors of 2 and 0.5.
 
For the \Zgamma process, the modelling uncertainty is derived by replacing the nominal sample with the alternative \MADGRAPH sample, interfaced to \PYTHIAV{8}. The fit is repeated with the alternative MC signal sample and then unfolded using the response matrix, signal efficiency and fake fraction from this alternative signal sample. Uncertainties in the signal efficiency and response matrix are already covered by experimental systematic errors outlined earlier.
 
The systematic uncertainty due to the dependence of the unfolding on the prior signal distribution, as obtained from MC simulations, is evaluated through a data-driven `closure test'. The simulated signal sample is reweighted at particle level such that the distribution of the fully simulated reconstructed jet mass more closely matches the observed data. Pseudo-data from the reweighted signal MC sample are then unfolded using the response matrix from the original unweighted signal MC sample, and the unfolded result is compared with the reweighted particle-level distribution. Differences observed in this comparison are taken as systematic uncertainties in the unfolding, and are referred to as unfolding non-closure uncertainties in the following. The uncertainty due to the dependence on the number of unfolding iteration steps is negligible. The statistical uncertainties in the signal MC sample, used to build the response matrix, and background templates are also considered.
 
A bootstrapping procedure~\cite{bootstrap} is used to ensure that the systematic uncertainties are statistically significant. For each systematic uncertainty considered, pseudo-experiments are constructed from the data or MC simulation by assigning each event a weight taken from a Poisson distribution with unit mean. The statistical uncertainty in the systematic variation is taken as the RMS across the pseudo-experiments. The jet mass distribution for each of the systematic variations is then rebinned until a target significance of $1.5$ standard deviations is achieved.
 
The dominant systematic uncertainties on the integrated fiducial cross-section measurements arise from the uncertainties in the fit, the signal modelling, the data-driven background estimate, the jet mass and energy scales, and the jet mass resolution. The uncertainty in the \pileup modelling in MC simulation is found to be negligible.
 
As shown in \Tab{\ref{tab:SysTable_crossSection_smoothed}} the impact of the statistical uncertainties on the fiducial cross-section of the response matrix and the background template is significantly different for trimmed and soft-drops jets. This behaviour can be explained by the differences of the signal and background jet mass distributions between trimmed and soft-drop jets and their interplay with the smoothing of those uncertainty components. For trimmed jets, only the variations in the core of the jet mass distributions are statistically significant while for soft-drop jets, the signal distribution is significantly wider and thus the tails also contribute to the systematic uncertainty. Furthermore, the background jet mass distribution of soft-drop jets is not as steeply falling as for trimmed jets. This results in less susceptibility to statistical fluctuations around the signal jet mass peak which would otherwise be reduced by the rebinning procedure introduced above.
 
% The next lines are included from the .//tables/table-systematics.tex input file
\begin{table}[h]
\begin{center}
\begin{tabular}{lcc}
\toprule
\multirow{2}{*}{Source} & \multicolumn{2}{c}{Uncertainty [\%]} \\
\cline{2-3}
& \multicolumn{1}{c}{Trimmed jets} & \multicolumn{1}{c}{Soft-drop jets} \\
\midrule
Luminosity & 2.1 & 2.1  \\
Jet energy resolution & 0.4 & $< 0.1$\hspace*{0.35cm}  \\
Jet mass resolution & 5.1 & 6.0  \\
Jet energy and mass scale & 7.2 & 7.4  \\
$b$-tagging & 5.3 & 5.8  \\
Photon related & 1.3 & 1.2  \\
Muon related & 0.1 & $< 0.1$\hspace*{0.35cm}  \\
Photon trigger & 0.4 & 0.4  \\
Transfer factor: 0-tag vs 1-tag & 7.5 & 4.0  \\
Transfer factor: statistical & 2.9 & 1.5  \\
\ttgamma related & 1.7 & 2.8  \\
\Wgamma related & $< 0.1$\hspace*{0.35cm} & $< 0.1$\hspace*{0.35cm}  \\
\Zgamma modelling & 12\hspace*{0.45cm} & 15\hspace*{0.45cm} \\
Unfolding non-closure & 9.4 & 5.8  \\
Signal MC response: statistical & 3.9 & 6.0  \\
Background template: statistical & 5.9 & 13\hspace*{0.45cm}  \\
\midrule
Fit statistical uncertainty & 30\hspace*{0.45cm} & 39\hspace*{0.45cm}  \\
\midrule
Total uncertainty & 37\hspace*{0.45cm} & 46\hspace*{0.45cm}  \\
\bottomrule
\end{tabular}
\caption{The uncertainties in the integrated fiducial cross-section measurement from data in the signal region for trimmed and soft-drop jets. Multiple independent components are combined into groups of systematic uncertainties.}
\label{tab:SysTable_crossSection_smoothed}
\end{center}
\end{table}
 
% End of text imported from the .//tables/table-systematics.tex input file
% End of text imported from the .//sections/sec-systematics.tex input file

\section{Results}
\label{sec:results}
% The next lines are included from the .//sections/sec-results.tex input file
Results of the measurement of the jet mass distribution in \Zbbgamma events are reported in the following three subsections; fit results and the calculation of the significance of the signal above the background, the unfolded fiducial cross-section measurement using the full measured jet mass spectrum, and the unfolded differential spectrum of the jet mass itself.

\subsection{Fit results and significance estimate}
\label{subsec:fitresults}
 
The signal yield is extracted by simultaneously fitting the signal and the background templates described in \Sect{\ref{sec:background}} to the observed \Zbb candidate jet mass ($m^{\Zjet}$) distribution. A binned maximum-likelihood fit is performed in the mass range between 30 and 160~\GeV\ using a bin width of 10~\GeV. The upper mass bound is chosen to exclude the mass region near the top quark mass while the lower mass bound is chosen to exclude the region of jet mass for which the uncertainty in the calibration is large and to protect against collinear singularities, as discussed in Section~\ref{sec:reconstruction}. The result of the fit to the reconstructed \Zbb candidate jet mass distribution is shown in \Figs{\ref{fig:data_mc_trimmed}}{\ref{fig:data_mc_softdrop}} for trimmed and soft-drop jets along with their corresponding background-subtracted data distributions in \Figs{\ref{fig:bkg_subtracted_trimmed}}{\ref{fig:bkg_subtracted_softdrop}}. The fitted signal yield is $215 \pm 61$ events for trimmed jets and $167 \pm 73$ events when using soft-drop jets as shown in \Tab{\ref{tab:eventyield_SR}}.

% The next lines are included from the .//tables/table-eventyield-SR.tex input file
 
\begin{table}[h]
\begin{center}

\small \begin{tabular}{l r@{\,\( \pm \)\,}r  r@{\,\( \pm \)\,}r}
 
\toprule
Process        & \multicolumn{2}{c}{Trimmed jets} & \multicolumn{2}{c}{Soft-drop jets}  \\
\midrule
 
\Zgamma & 215 & 61  & 167 & 73  \\
\gammajets & 4180 &  90 &  4630 & 100  \\
\Wgamma & 39 & 8 &  37 & 8  \\
\ttgamma & 39 & 12 &  40 & 12  \\
\midrule
Total & 4480 & 110 & 4870 & 120  \\

\midrule
Data           & \multicolumn{2}{c}{4475}  & \multicolumn{2}{c}{4874} \\
\bottomrule
\end{tabular}
\caption{The number of data events observed in the signal region, along with the composition of these events after the fit in the $30 < m^{\Zjet} < 160$~\GeV\ mass range. Numbers are presented for trimmed and soft-drop jets at reconstruction level. Statistical and systematic uncertainties are added in quadrature. Systematic uncertainties are described in Section~\ref{sec:systematics}.}
\label{tab:eventyield_SR}
\end{center}
\end{table}
 
% End of text imported from the .//tables/table-eventyield-SR.tex input file

\begin{figure}[htbp!]
\begin{center}
\subfloat[]{\includegraphics[width = 0.49\textwidth]{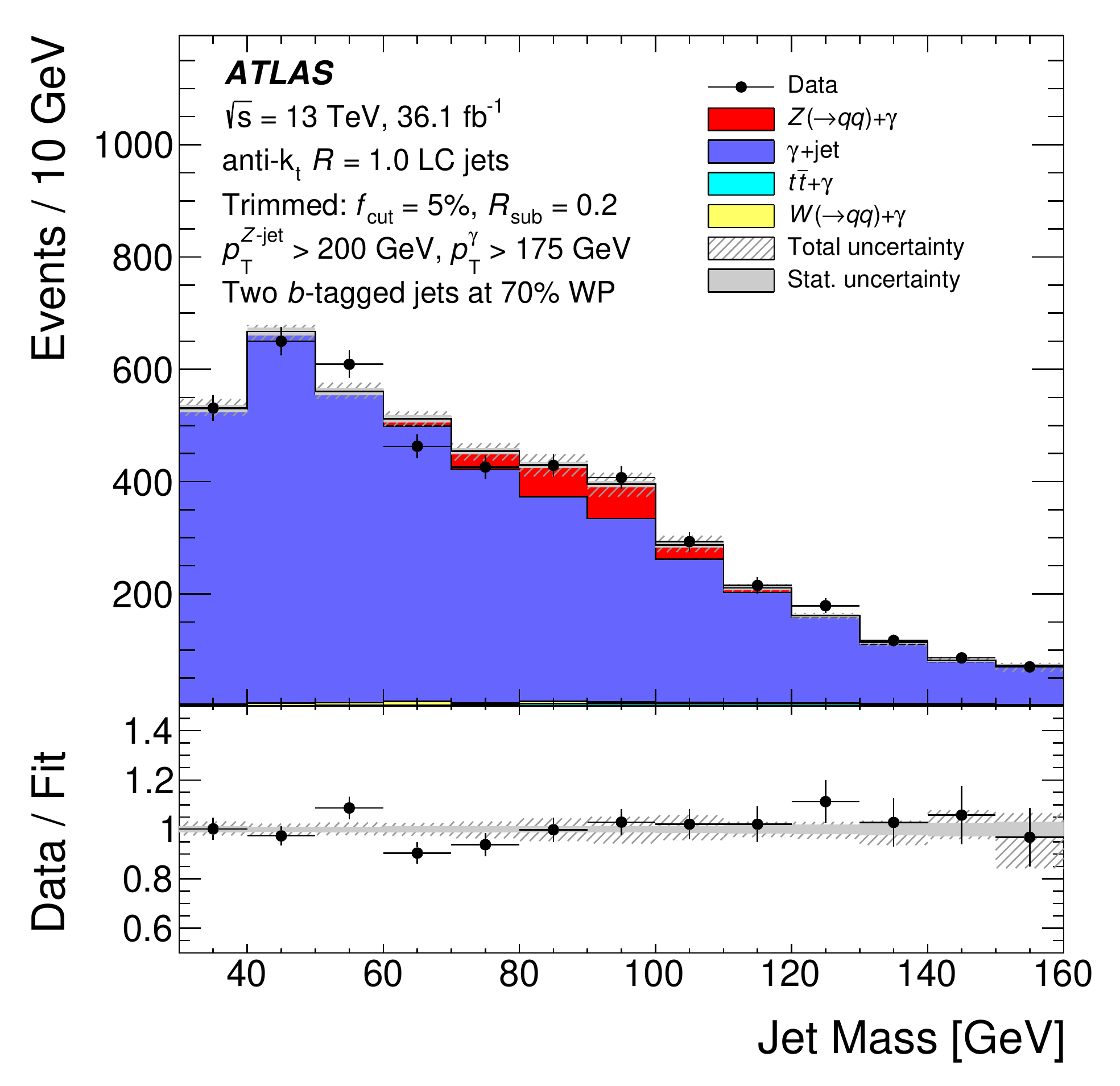}\label{fig:data_mc_trimmed}}
\subfloat[]{\includegraphics[width = 0.49\textwidth]{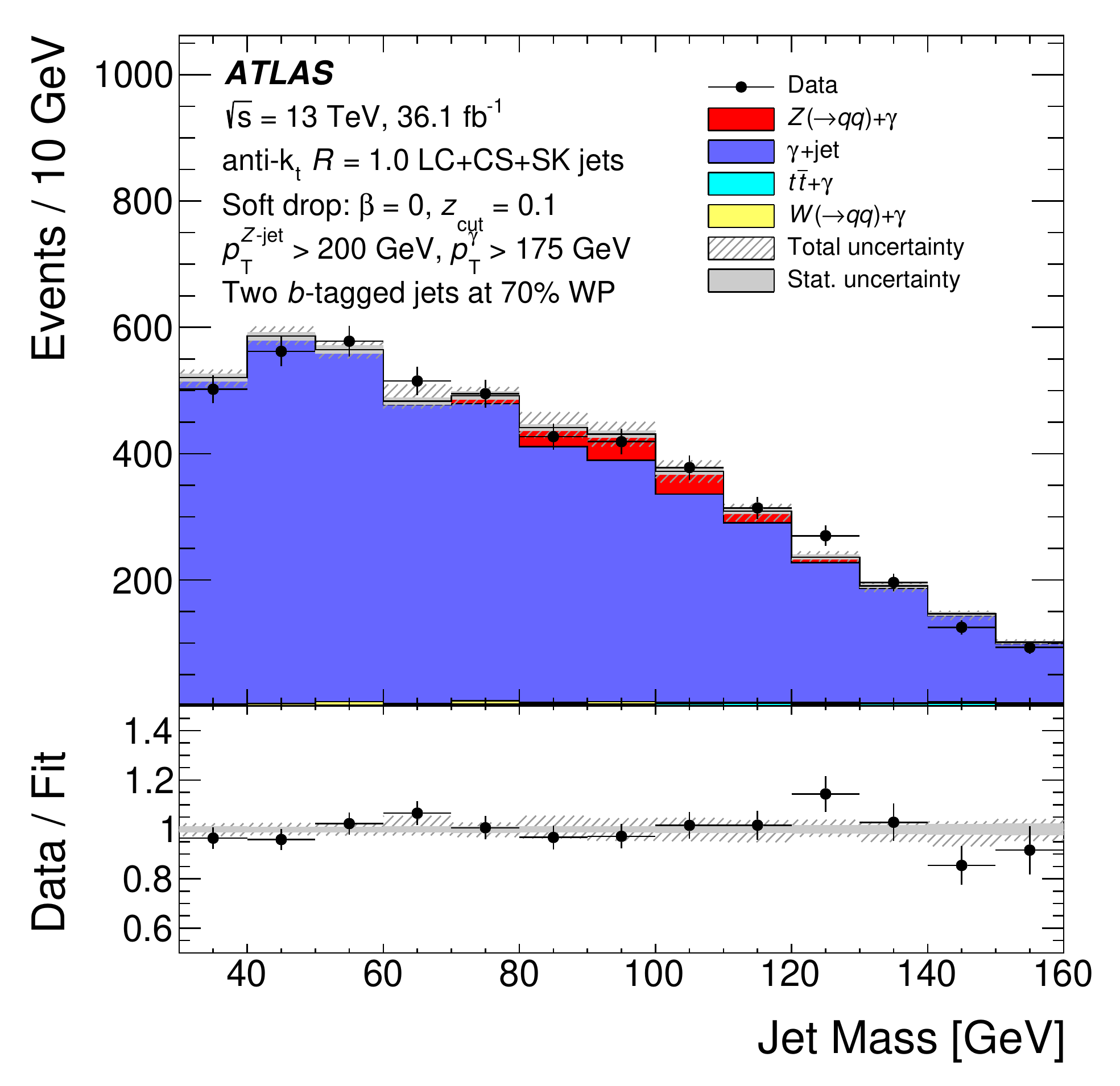}\label{fig:data_mc_softdrop}}
 
\subfloat[]{\includegraphics[width = 0.49\textwidth]{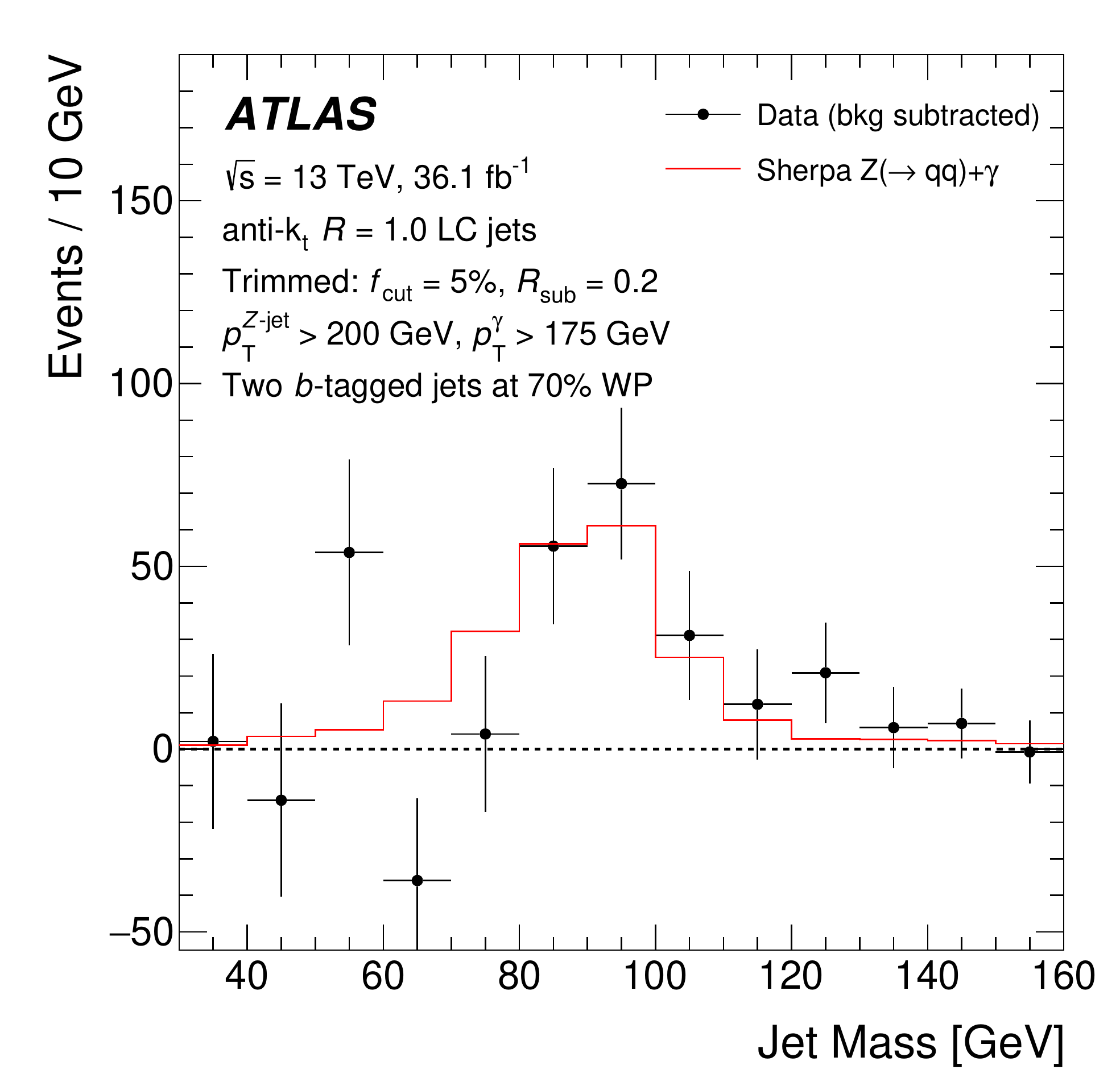}\label{fig:bkg_subtracted_trimmed}}
\subfloat[]{\includegraphics[width = 0.49\textwidth]{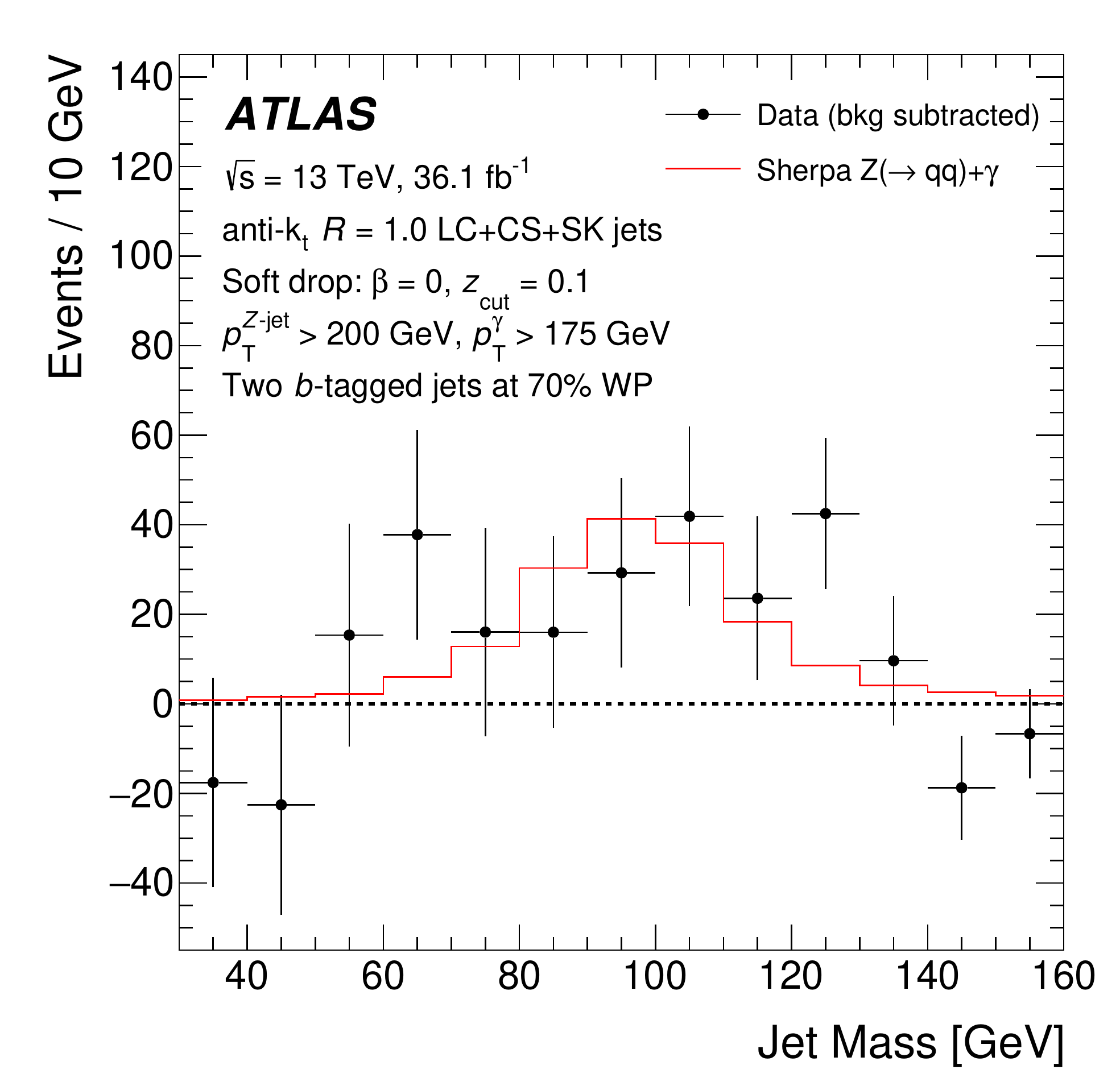}\label{fig:bkg_subtracted_softdrop}}
\end{center}
\caption{The reconstructed jet mass distribution in the signal region (a, b) after fitting the \SHERPAV{2.1.1} signal model and background templates to the data and (c, d) the corresponding background-subtracted distributions for (a, c) trimmed and (b, d) soft-drop jets. The ratio of data to the fitted signal plus background is shown at the bottom in Figures (a) and (b). The background-subtracted jet mass distributions in data are compared with the reconstructed signal jet mass distributions in the nominal Monte Carlo simulation. The error bars on the background-subtracted data distribution are statistical only. The \SHERPA signal model and background template are scaled to fit the data as described in Section~\ref{sec:background}.}
\label{fig:sig_fitResult_usingdatatemplate_jetmass}
\end{figure}
 
The 13 bins of the \Zbb candidate jet mass distribution are combined in a profile likelihood fit~\cite{Cranmer:1456844} to extract the expected and observed significances. Systematic uncertainties are included in the fit as nuisance parameters and are assumed to be gaussian distributed. The expected and observed significances of the Standard Model prediction fitted to the observed data for the \Zbbgamma production are summarised in \Tab{\ref{tab:Significance}}. For each jet definition, the observed significance is consistent with the expectation.  Differences in the significance between the two jet definitions are related to the differences in both the jet mass resolution and the fiducial cross-sections between trimmed and soft-drop jets, which affect the signal and background yields in the 30--160 GeV mass window.

% The next lines are included from the .//tables/table-significances.tex input file
\begin{table}[h]
\begin{center}
\begin{tabular}{  l c c }
\toprule
& Trimmed jets & Soft-drop jets \\
\midrule
Expected significance & $3.8$ & $2.7$ \\
\midrule
Observed significance & $3.9$ & $2.7$ \\
\bottomrule
\end{tabular}
\caption{Expected and observed significance values (in numbers of standard deviations) for trimmed and soft-drop jets for a mass range between 30 and 160~\GeV. }
\label{tab:Significance}
\end{center}
\end{table}
% End of text imported from the .//tables/table-significances.tex input file
 
\subsection{Integrated fiducial cross-section measurement}
\label{subsec:fiducial}

The measured integrated fiducial cross-sections in the boosted (high $\pt$) \Zbb region are listed in \Tab{\ref{tab:SignalRegion_crossSection}}. The integrated fiducial cross-section is extracted from the ratio of the unfolded yield of signal events and the total integrated luminosity. The measurements are compared with the \SHERPAV{2.1.1} and \MadgraphPythiaV LO predictions described in \Sect{\ref{sec:dataMC}}. \MadgraphPythiaV predicts around 30\% fewer events than the samples generated with \Sherpa. Within the current uncertainties on the measurement, both predictions are consistent with the measured cross-sections for soft-drop jets. For trimmed jets, larger differences can be observed between the \MadgraphPythiaV prediction and the measured cross-section. The uncertainties in the measured integrated fiducial cross-section results are summarised in \Tab{\ref{tab:SysTable_crossSection_smoothed}}. The dominant source of systematic uncertainty is the fit uncertainty for both jet definitions. The uncertainty in the normalisation of the background template has a large impact on the cross-section measurement because of the order of magnitude difference between the estimated numbers of signal and background events.

% The next lines are included from the .//tables/table-crosssections.tex input file
\begin{table}[h]
\begin{center}
\begin{tabular}{  l l S[separate-uncertainty=true,table-figures-uncertainty=1] l }
\toprule
Jet definition       & \multicolumn{3}{l}{$\sigma$ (\Zbbgamma, $\ptZjet > 200$~\GeV, $\ptg > 175$~\GeV, $30 < \mZjet < 160~\GeV$) [fb]}\\
\midrule
\multirow{2}{*}{Trimmed jets}  &  Data      & 17.0 \pm 5.0 &\hspace{-10px}(stat.) $\pm$ 3.6 (syst.) \\
&  \Sherpa $Z\gamma$                    prediction   	& 13.4 \pm 0.2 &\hspace{-10px}(stat.) \\
&  \MadgraphPythiaV $Z\gamma$ prediction            &  9.1 \pm 0.1 &\hspace{-10px}(stat.) \\
\midrule
\multirow{2}{*}{Soft-drop jets}  &  Data      & 12.5 \pm 4.9 &\hspace{-10px}(stat.) $\pm$ 3.1 (syst.) \\
&  \Sherpa $Z\gamma$                    prediction      & 15.4 \pm 0.1 &\hspace{-10px}(stat.) \\
&  \MadgraphPythiaV $Z\gamma$ prediction              & 10.2 \pm 0.1 &\hspace{-10px}(stat.) \\
\bottomrule
\end{tabular}
\caption{Measured and predicted cross-section for \Zbbgamma in the $\ptZjet > 200~\GeV$, $30 < \mZjet < 160~\GeV$ and $\ptg > 175$~\GeV\ region.}
\label{tab:SignalRegion_crossSection}
\end{center}
\end{table}
 
% End of text imported from the .//tables/table-crosssections.tex input file

\subsection{Differential fiducial cross-section measurement}
\label{subsec:differential}
 
The differential fiducial cross-section of \Zbbgamma production as a function of the \Zbb jet mass, obtained from the unfolded data in the signal region, is shown in \Fig{\ref{fig:sig_unfoldedResult_usingdatatemplate_jetmass}} for trimmed and soft-drop jets. As a comparison, the prediction from \SHERPAV{2.1.1} at LO is also shown. 
Statistical uncertainties are significant for the differential fiducial cross-section measurement in the tails of the jet mass distribution.
 
\begin{figure}[htbp!]
\begin{center}
 
\subfloat[]{\includegraphics[width = 0.6\textwidth]{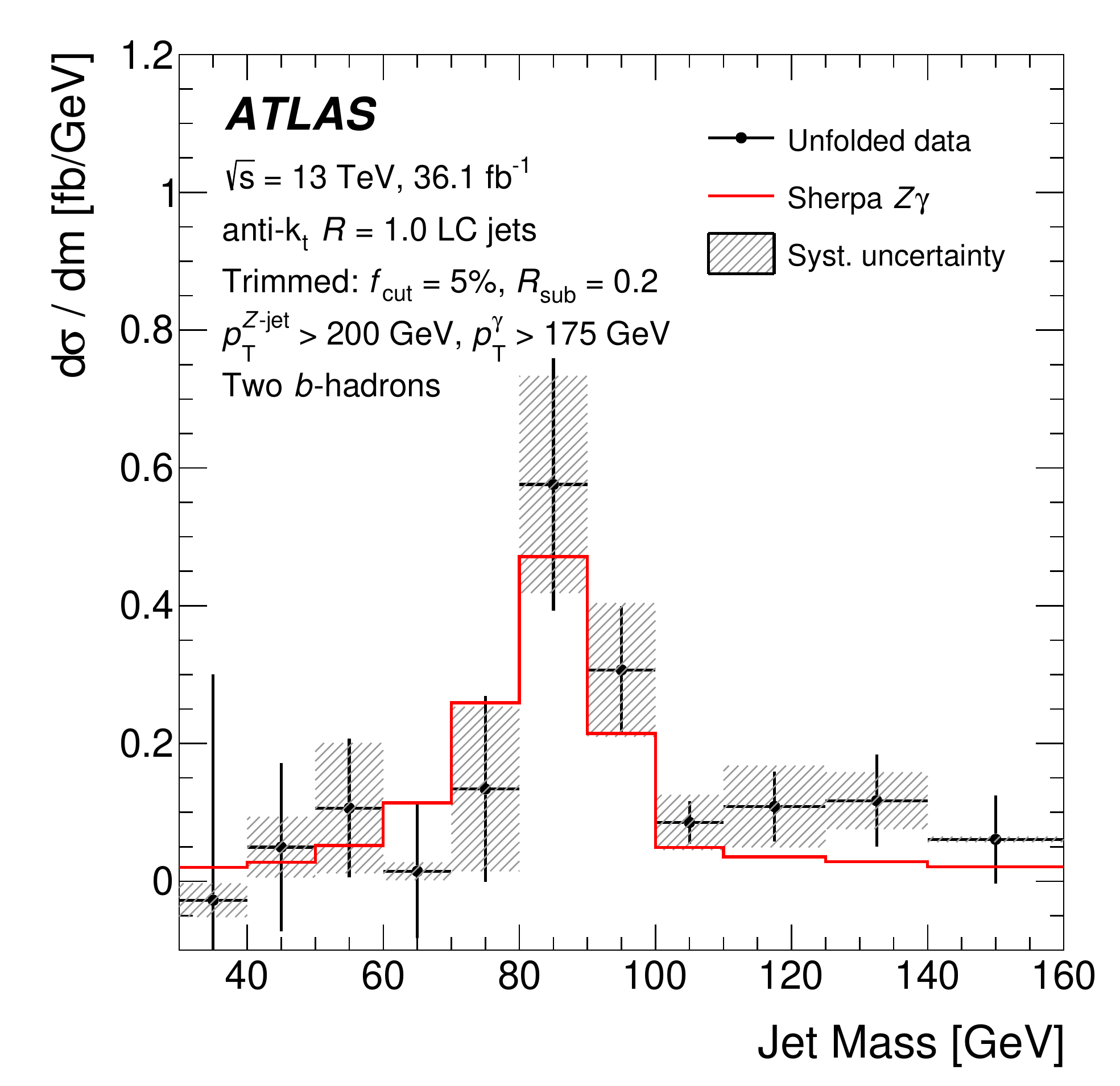}\label{fig:unfolded_trimmed}}
 
\subfloat[]{\includegraphics[width = 0.6\textwidth]{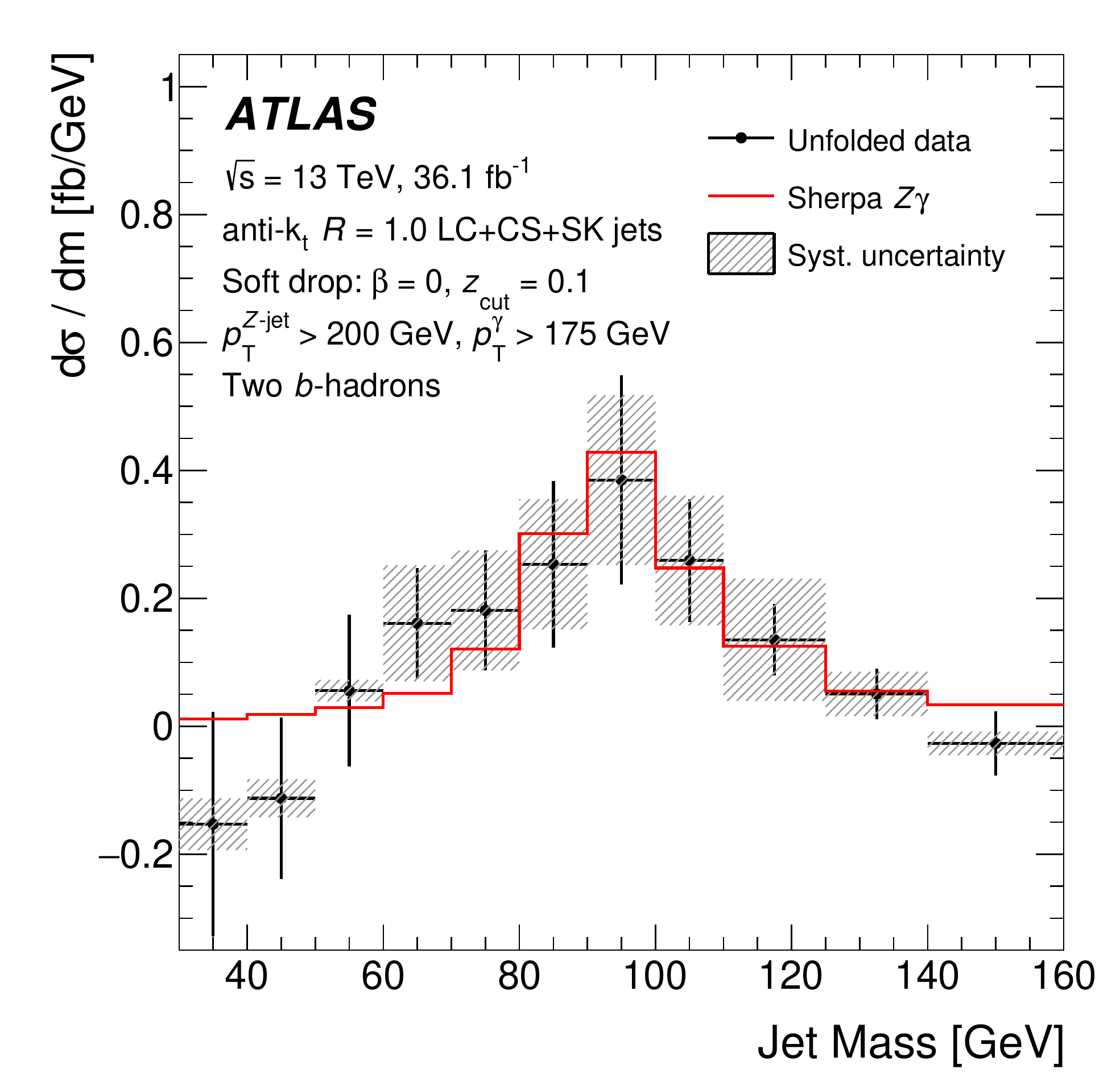}\label{fig:unfolded_SD}}
 
\end{center}
\caption{Unfolded distribution of the \Zbb candidate jet mass from background-subtracted data in the signal region along with the predictions from \SHERPAV{2.1.1}. Results for (a) trimmed and (b) soft-drop jets are shown. The error bars correspond to the statistical uncertainty while the hatched band corresponds to the total systematic uncertainty in the measurement. The statistical uncertainty in the \SHERPA signal is negligible and not shown in the figure.} 
\label{fig:sig_unfoldedResult_usingdatatemplate_jetmass}
\end{figure}

% End of text imported from the .//sections/sec-results.tex input file

\FloatBarrier

\section{Conclusion}
\label{sec:conclusion}
% The next lines are included from the .//sections/sec-conclusion.tex input file
The fully unfolded differential jet mass spectrum for the high-\pt \Zbb signal using the $Z\gamma$ final state and the fiducial production cross-section are measured in \intlumi of \pp collisions at \sqsthirt recorded in 2015 and 2016 by the ATLAS detector. The high-\pt \Zbb signal is reconstructed using large-$R$ jets and jet substructure techniques, including double subjet $b$-tagging. Two different grooming algorithms are used in this analysis: trimming and soft drop. These grooming algorithms exhibit differences in the measured shapes of the jet mass spectra and the resulting precision of those measurements. The soft drop jet mass spectrum is observed to be broader and more symmetric than that of trimmed jets. The precision of the fiducial cross-section measured is slightly better for trimmed jets. Within the fiducial regions defined for each jet definition as $\ptZjet > 200~\GeV$, $30 < \mZjet < 160~\GeV$ and $\ptg > 175$~\GeV\ , the production cross-sections are measured to be:
 
\begin{center}
 
$\sigma_{\Zbbgamma}^{\text{trimming}}=17.0 \pm 5.0\ \text{(stat.)} \pm 3.6\ \text{(syst.)}$ fb,\\
$\sigma_{\Zbbgamma}^{\text{soft drop}}=12.5 \pm 4.9\ \text{(stat.)} \pm 3.1\ \text{(syst.)}$ fb.
 
\end{center}
 
The integrated fiducial cross-section measurements and the differential cross-section of the jet mass of the boosted \Zbb decay are found to be in agreement with the LO predictions from \Sherpa and \MadgraphPythiaV (integrated), and from \Sherpa (differential), respectively, within the current statistical and systematic uncertainties. Future measurements of resonant $\bbbar$ decays will need to account for the observed experimental features of these different grooming algorithms in concert with the availability of precision theoretical calculations of the jet mass spectra, or lack thereof.
% End of text imported from the .//sections/sec-conclusion.tex input file

\section*{Acknowledgements}
 
% The next lines are included from the .//acknowledgements/Acknowledgements.tex input file

We thank CERN for the very successful operation of the LHC, as well as the
support staff from our institutions without whom ATLAS could not be
operated efficiently.
 
We acknowledge the support of ANPCyT, Argentina; YerPhI, Armenia; ARC, Australia; BMWFW and FWF, Austria; ANAS, Azerbaijan; SSTC, Belarus; CNPq and FAPESP, Brazil; NSERC, NRC and CFI, Canada; CERN; ANID, Chile; CAS, MOST and NSFC, China; COLCIENCIAS, Colombia; MSMT CR, MPO CR and VSC CR, Czech Republic; DNRF and DNSRC, Denmark; IN2P3-CNRS and CEA-DRF/IRFU, France; SRNSFG, Georgia; BMBF, HGF and MPG, Germany; GSRT, Greece; RGC and Hong Kong SAR, China; ISF and Benoziyo Center, Israel; INFN, Italy; MEXT and JSPS, Japan; CNRST, Morocco; NWO, Netherlands; RCN, Norway; MNiSW and NCN, Poland; FCT, Portugal; MNE/IFA, Romania; JINR; MES of Russia and NRC KI, Russian Federation; MESTD, Serbia; MSSR, Slovakia; ARRS and MIZ\v{S}, Slovenia; DST/NRF, South Africa; MICINN, Spain; SRC and Wallenberg Foundation, Sweden; SERI, SNSF and Cantons of Bern and Geneva, Switzerland; MOST, Taiwan; TAEK, Turkey; STFC, United Kingdom; DOE and NSF, United States of America. In addition, individual groups and members have received support from BCKDF, CANARIE, Compute Canada, CRC and IVADO, Canada; Beijing Municipal Science \& Technology Commission, China; COST, ERC, ERDF, Horizon 2020 and Marie Sk{\l}odowska-Curie Actions, European Union; Investissements d'Avenir Labex, Investissements d'Avenir Idex and ANR, France; DFG and AvH Foundation, Germany; Herakleitos, Thales and Aristeia programmes co-financed by EU-ESF and the Greek NSRF, Greece; BSF-NSF and GIF, Israel; La Caixa Banking Foundation, CERCA Programme Generalitat de Catalunya and PROMETEO and GenT Programmes Generalitat Valenciana, Spain; G\"{o}ran Gustafssons Stiftelse, Sweden; The Royal Society and Leverhulme Trust, United Kingdom.
 
The crucial computing support from all WLCG partners is acknowledged gratefully, in particular from CERN, the ATLAS Tier-1 facilities at TRIUMF (Canada), NDGF (Denmark, Norway, Sweden), CC-IN2P3 (France), KIT/GridKA (Germany), INFN-CNAF (Italy), NL-T1 (Netherlands), PIC (Spain), ASGC (Taiwan), RAL (UK) and BNL (USA), the Tier-2 facilities worldwide and large non-WLCG resource providers. Major contributors of computing resources are listed in Ref.~\cite{ATL-SOFT-PUB-2020-001}.
 
% End of text imported from the .//acknowledgements/Acknowledgements.tex input file

\printbibliography

\clearpage % ATLAS Collaboration author list
% Reference date of STDM-2017-14 is 2019-02-27
% Author list last updated on date 27-NOV-20
% Data extracted on 27-Nov-2020 for paper reference STDM-2017-14
% at 2:28pm
 
\begin{flushleft}
\hypersetup{urlcolor=black}
{\Large The ATLAS Collaboration}

\bigskip

\AtlasOrcid[0000-0002-6665-4934]{G.~Aad}$^\textrm{\scriptsize 102}$,    
\AtlasOrcid[0000-0002-5888-2734]{B.~Abbott}$^\textrm{\scriptsize 129}$,    
\AtlasOrcid{D.C.~Abbott}$^\textrm{\scriptsize 103}$,    
\AtlasOrcid{O.~Abdinov}$^\textrm{\scriptsize 13,*}$,    
\AtlasOrcid[0000-0002-2788-3822]{A.~Abed~Abud}$^\textrm{\scriptsize 71a,71b}$,    
\AtlasOrcid[0000-0002-1002-1652]{K.~Abeling}$^\textrm{\scriptsize 53}$,    
\AtlasOrcid[0000-0002-2987-4006]{D.K.~Abhayasinghe}$^\textrm{\scriptsize 94}$,    
\AtlasOrcid[0000-0002-8496-9294]{S.H.~Abidi}$^\textrm{\scriptsize 168}$,    
\AtlasOrcid[0000-0002-8279-9324]{O.S.~AbouZeid}$^\textrm{\scriptsize 40}$,    
\AtlasOrcid{N.L.~Abraham}$^\textrm{\scriptsize 157}$
% ...
\bigskip
\\

% ...
$^{13}$Institute of Physics, Azerbaijan Academy of Sciences, Baku; Azerbaijan.\\
% ...
$^{40}$Niels Bohr Institute, University of Copenhagen, Copenhagen; Denmark.\\
% ...
$^{53}$II. Physikalisches Institut, Georg-August-Universit\"{a}t G\"ottingen, G\"ottingen; Germany.\\
% ...
$^{71}$$^{(a)}$INFN Sezione di Pavia;$^{(b)}$Dipartimento di Fisica, Universit\`a di Pavia, Pavia; Italy.\\
% ...
$^{94}$Department of Physics, Royal Holloway University of London, Egham; United Kingdom.\\
% ...
$^{102}$CPPM, Aix-Marseille Universit\'e, CNRS/IN2P3, Marseille; France.\\
$^{103}$Department of Physics, University of Massachusetts, Amherst MA; United States of America.\\
% ...
$^{129}$Homer L. Dodge Department of Physics and Astronomy, University of Oklahoma, Norman OK; United States of America.\\
% ...
$^{157}$Department of Physics and Astronomy, University of Sussex, Brighton; United Kingdom.\\
% ...
$^{168}$Department of Physics, University of Toronto, Toronto ON; Canada.\\
% ...
$^{a}$ Also at Borough of Manhattan Community College, City University of New York, New York NY; United States of America.\\
$^{b}$ Also at Centre for High Performance Computing, CSIR Campus, Rosebank, Cape Town; South Africa.\\
% ...
$^{*}$ Deceased

\end{flushleft}

% Created with Glance <Atlas.Glance@cern.ch>

\end{document}